\documentstyle[prb,aps,twocolumn,fleqn,eqsecnum]{revtex}

\begin{document}
\draft
\twocolumn[\hsize\textwidth\columnwidth\hsize\csname@twocolumnfalse%
\endcsname
\title{Trace and antitrace maps for aperiodic sequences, 
       their extensions and applications}
\author{Xiaoguang Wang,$^{1,2,3}$ Uwe Grimm,$^{1,4}$
and Michael Schreiber$^{1}$\vspace{1ex}}

\address{$^{1}$Institut f\"{u}r Physik,
               Technische Universit\"{a}t,
               09107 Chemnitz, Germany}
\address{$^{2}$Institute of Physics and Astronomy, 
               Aarhus University, 8000 Aarhus, Denmark}
\address{$^{3}$Laboratory of Optical Physics, 
               Institute of Physics, 
               Chinese Academy of Sciences, \\
               Beijing 100080, People's Republic of China}
\address{$^{4}$Applied Mathematics Department, 
               Faculty of Mathematics and Computing,\\
               The Open University,
               Milton Keynes MK7 6AA, U.K.}
\date{September 5, 2000}
\maketitle

\begin{abstract}%
We study aperiodic systems based on substitution rules by means of a
transfer-matrix approach.  In addition to the well-known trace map, we
investigate the so-called ``antitrace'' map, which is the
corresponding map for the difference of the off-diagonal elements of
the $\!2\times 2\!$ transfer matrix.  The antitrace maps are obtained
for various binary, ternary, and quaternary aperiodic sequences, such
as the Fibonacci, Thue-Morse, period-doubling, Rudin-Shapiro
sequences, and certain generalizations. For arbitrary substitution
rules, we show that not only trace maps, but also antitrace maps
exist.  The dimension of the our antitrace map is $r(r+1)/2$, where
$r$ denotes the number of basic letters in the aperiodic
sequence. Analogous maps for specific matrix elements of the transfer
matrix can also be constructed, but the maps for the off-diagonal
elements and for the difference of the diagonal elements coincide with
the antitrace map.  Thus, from the trace and antitrace map, we can
determine any physical quantity related to the global transfer matrix
of the system. As examples, we employ these dynamical maps to compute
the transmission coefficients for optical multilayers, harmonic
chains, and electronic systems.
\end{abstract}


\vspace*{5ex}

]
\narrowtext

\section{Introduction}
\label{sec:intro}

The trace-map technique, first introduced in 1983,\cite{Kohmoto83} has
proven to be a powerful tool to investigate the electronic spectrum of
various aperiodic systems, such as the Fibonacci sequence
(FS),\cite{Kohmoto83} the Thue-Morse sequence (TMS),\cite{Axel89} and
the period-doubling sequence.\cite{Luck89} It has also been applied to
investigate other physical systems, for instance, kicked two-level
systems,\cite{Kicked,Baake} and classical and quantum spin
systems.\cite{Spin} The technique was extended to study aperiodic
systems in combination with the real-space renormalization-group
technique.\cite{Ghosh98} Recently, trace maps have been used to
evaluate localization properties in a FS tight-binding
model.\cite{Naumis99}

This technique was transferred to the field of optics in order to see
the scaling of the light transmission coefficient through a Fibonacci
dielectric multilayer.\cite{Optics} Now, we consider light that is
vertically transmitted through a Fibonacci multilayer of two materials
$a$ and $b$ which is sandwiched by two media of type $a$.  The FS is
constructed by the substitution rule $b\rightarrow a$, $a\rightarrow
ab$.  The corresponding transfer matrices $A_l$ are written
as\cite{Optics}
\begin{eqnarray}
A_1&=&P_{ab}{P}_bP_{ba},\nonumber\\
A_2&=&{P}_a,\nonumber\\
A_{l+1}&=&A_lA_{l-1},
\label{eq:opt}
\end{eqnarray}
where ${P}_{ab} ({P}_{ba})$ stands for the propagation matrix from
layer $a$ to $b$ ($b$ to $a$) and ${P}_{a}$ is the propagation matrix
through the single layer $a$. They are given by\cite{Optics}
\begin{eqnarray}
P_{ab}&=&P_{ba}^{-1}=\left(\matrix{1&0\cr 0&n_a/n_{b}}\right),\nonumber\\
P_{a}&=&\left(\matrix{\cos\delta_{a}&-\sin\delta_{a}\cr 
\sin\delta_{a}&\cos\delta_{a}}\right),
\label{eq:phase}
\end{eqnarray} 
where $\delta_{a}=kn_{a}d_{a}$, $n_{a}$ is the refraction index of
material $a$, $d_{a}$ denotes the thickness of the layers, and $k$ is
the wave number in vacuum.  The quantity $\delta_{a}$ is the phase
difference between the ends of a layer.  For material $b$, the
quantities $P_b$, $\delta_b$, $n_b$, and $d_b$ are defined
analogously.

The transmission coefficient is given by\cite{Optics}
\begin{equation}
t_l=\frac{4}{|A_l|^2+2},
\end{equation}
where $|A_l|^2$ is the sum of squares of the four elements of
$A_l$. Since the transfer matrix is unimodular, we can express the
transmission coefficient as
\begin{equation}
t_l=\frac{4}{x_l^2+y_l^2},
\label{eq:t}
\end{equation}
where $x_l$ and $y_l$ denote the trace and antitrace of the transfer
matrix $A_l$, respectively. Here, the so-called ``antitrace" of a
$2\!\times\! 2$ matrix
\begin{equation}
A = \left(\matrix{A_{11} & A_{12}\cr
                  A_{21} & A_{22}}\right)
\end{equation}
is defined as $y_A=A_{21}-A_{12}$, which follows the notion of
Ref.\onlinecite{Dulea90}. {}From Eq.~(\ref{eq:t}), we see that the
transmission coefficient is completely determined by the trace and the
antitrace, i.e., a complete description of the light transmission
through general aperiodic multilayers requires both the trace and
the antitrace map.\cite{Dulea90}

Now, we consider a different system, namely a harmonic chain composed
of two kinds of masses, $m_a$ and $m_b$, which are arranged according
to the FS, and are coupled by two kinds of springs, $K_{aa}$ and
$K_{ab}=K_{ba}$.  Making use of the transfer-matrix formalism, the
equation of motion is\cite{Macia99,MacDA}
\begin{eqnarray}
{u_{n+1} \choose u_n}&=&\left(\matrix {\frac{a_n}{K_{n,n+1}}&
-\frac{K_{n,n-1}}{K_{n,n+1}}\cr 1&0} \right){u_n\choose u_{n-1}}\nonumber\\
&=&{ P}_n{u_n\choose u_{n-1}},
\end{eqnarray}
where $u_n$ is the displacement of the $n$th atom from its equilibrium
position, $K_{n,n\pm 1}$ denotes the strength of the harmonic coupling
between neighboring atoms, $a_n=K_{n,n-1}+K_{n,n+1}-m_n\omega^2$, and
$\omega$ is the vibration frequency.

{}From the corresponding global transfer matrix
$A_l=\prod_{n=N_l}^{1}P_n$, the transmission coefficient $t_l$ and the
Lyapunov exponent $\Gamma_l$ are given by\cite{Macia99,MacDA}
\begin{eqnarray}
t_l&=&\frac{4\sin^2k}{(z_l\cos k-y_l)^2+x_l^2\sin^2k}, \label{eq:ttt}\\
\Gamma_l&=&
\frac{1}{N_l}\ln(|A_l|^2) =
\frac{1}{N_l}\ln(x_l^2+y_l^2-2),\label{eq:Gamma}
\end{eqnarray}
where $z_l=(A_l)_{11}-(A_l)_{22}$, $N_l$ denotes the number of atoms in
the chain, and $\cos k=m_a/(2K_{aa})$. 

As our third example of physical systems based on aperiodic
substitution sequences, we consider the transmission in electronic
systems. This is closely related to the harmonic chain considered
above. The Schr\"{o}dinger equation for a one-dimensional
tight-binding model with nearest-neighbor hopping can be written in
matrix form as follows\cite{KraSch,Macia}
\begin{eqnarray}
{\phi_{n+1} \choose \phi_n}&=&
\left(\matrix {{E-\epsilon_n\over t_{n,n+1}}&
-{t_{n,n-1}\over t_{n,n+1}}\cr 1&0} \right)
{\phi_n\choose \phi_{n-1}}\nonumber\\  
&=&M_n{\phi_n\choose \phi_{n-1}},
\label{eq:Schroedinger}
\end{eqnarray}
where $\phi_n$ denotes the amplitude of the wave function in the
Wannier representation, $E$ the corresponding energy, $\epsilon_n$ the
on-site energy at site $n$, and $t_{n,n\pm 1}$ the hopping matrix
element between two neighboring sites. $M_n$ is the local transfer
matrix associated with site $n$. The transmission coefficient is given
by\cite{MacDA,Macia}
\begin{eqnarray}
t_l&=&\frac{4-E^2}{(z_lE/2-y_l)^2+x_l^2(1-E^2/4)}, \label{eq:tttttt}
\end{eqnarray}
where the quantities $x_l$, $y_l$, and $z_l$ are again related to the
global transfer matrix of the chain, i.e., the product
$A_l=\prod_{n=N_l}^{1}M_n$ of the local transfer matrices along the
chain. Note the similarity to Eq.~(\ref{eq:ttt}). The corresponding
Lyapunov exponent $\Gamma_l$ is given by the same expression
(\ref{eq:Gamma}).

For the latter two systems, the Lyapunov exponent is completely
determined by the trace and the antitrace; however, we need to know
$z_l$ to calculate the transmission coefficient.  Fortunately, it
turns out that the maps for $z_l$ and $y_l$ are the same, as will be
shown in Sec.~\ref{sec:mem}. Therefore, the trace and antitrace map
are sufficient to determine the transmission coefficient and the
Lyapunov exponent. Thus it is desirable to construct antitrace maps
for various aperiodic sequences, which is the motivation of the work
presented here.

The paper is organized as follows. In Sec.~\ref{sec:tam}, we give the
antitrace maps for various classes of aperiodic sequences, including
the FS, the TMS, the periodic-doubling sequence, and certain
generalizations.  The extension to arbitrary substitution rules and to
maps for matrix elements are discussed in Sec.~\ref{sec:ass} and in
Sec.~\ref{sec:mem}, respectively. It is shown that the antitrace maps
and the maps for matrix elements exist for arbitrary substitution
rules and that the maps for non-diagonal elements and for the
difference of the diagonal elements coincide with the antitrace maps.
Applications to the computation of transmission coefficients and
Lyapunov exponents in different aperiodic systems are investigated in
Sec.~\ref{sec:app}. Finally, in Sec.~\ref{sec:con}, we conclude.

\section{Trace and antitrace maps for two-letter sequences}
\label{sec:tam}

We now proceed with the derivation of the antitrace maps of various
classes of aperiodic sequences.  We also include the corresponding
trace maps for reasons which will become clear later. In this part, we
make ample use of several relations for unimodular
matrices. Therefore, we append a compilation of these relations in
Appendix~\ref{sec:rum}.

\subsection{Generalized Fibonacci sequences}

There are many kinds of generalized FSs.  Here, we study two-letter
sequences $\text{FS}(m,n)$ that can be generated by the inflation
scheme\cite{Kolar90,GFS}
\begin{equation}
S_0=b,\quad S_1=a,\quad S_{l+1}=S^m_lS^n_{l-1}
\label{eq:fs}
\end{equation} 
with arbitrary positive integers $m$ and $n$, where $\text{FS}(1,1)$
corresponds to the well-known standard FS.  Equivalently, they can
also be generated by the substitution rule
\begin{equation}
b\rightarrow a, \quad
a\rightarrow a^mb^n.
\end{equation}
The total number of letters $a$ and $b$ in the word $S_{l}$ is denoted
by $F_l$ and satisfies the recursion relation
\begin{equation}
F_{l+1}=mF_l+nF_{l-1},\quad F_0=F_1=1.
\end{equation}
In the limit of an infinite sequence, the ratio of word lengths for
subsequent inflation steps is given by
\begin{equation}
\sigma=\lim_{l\rightarrow \infty}\frac{F_{l+1}}{F_{l}}=
\frac{m+\sqrt{m^2+4n}}{2}.
\end{equation}
Some values of $\sigma$ and commonly used terms for special cases of
so-called ``metallic means''\cite{Spinadel} are
\begin{eqnarray*}
\text{FS}(1,1):\quad & &
\makebox[15ex][l]{$\displaystyle\sigma_g=\frac{1+\sqrt{5}}{2}$}\quad 
\text{golden mean,}\\
\text{FS}(2,1):\quad & &
\makebox[15ex][l]{$\displaystyle\sigma_s= 1+\sqrt{2}$}\quad
\text{silver mean,}\\
\text{FS}(3,1):\quad & &
\makebox[15ex][l]{$\displaystyle\sigma_b=\frac{3+\sqrt{13}}{2}$}\quad
\text{bronze mean,}\\ 
\text{FS}(1,2):\quad & &
\makebox[15ex][l]{$\displaystyle\sigma_c= 2$}\quad
\text{copper mean,}\\
\text{FS}(1,3):\quad & &
\makebox[15ex][l]{$\displaystyle\sigma_n=\frac{1+\sqrt{13}}{2}$}\quad
\text{nickel mean}.
\end{eqnarray*}
It is known that the sequences $\text{FS}(m,n)$ with $n=1$ are
quasiperiodic and those with $n\ge 2$ are always aperiodic.

It is interesting to consider two further classes of generalized
FSs\cite{Fu97,XBYang99}
\begin{eqnarray}
b&\rightarrow& b^{m-1}a,\quad a\rightarrow b^{m-1}ab, 
\label{eq:fc1}\\
b&\rightarrow& b^{m-2}a,\quad a\rightarrow b^{m-2}ab^{m-2}ab. 
\label{eq:fc2}
\end{eqnarray}
The first class (\ref{eq:fc1}) consists of the so-called
Fibonacci-class sequences $\text{FC}(m)$\cite{Fu97,XBYang99}, the
second (\ref{eq:fc2}) occurs in the renormalization-group analysis of
the energy spectrum of $\text{FC}(m)$ chains.\cite{Fu97} It is easy to
check that the inflation schemes of these two generalized FSs are the
same as those for $\text{FS}(m,1)$, but they differ in the initial
words.  A natural further generalization of these sequences is given
by
\begin{eqnarray}
b\rightarrow b^{m-k}a,\quad a\rightarrow (b^{m-k}a)^kb,
\end{eqnarray}
which we denote as $\text{FC}(m,k)$. Here, $\text{FC}(m,1)$ and
$\text{FC}(m,2)$ correspond to the cases (\ref{eq:fc1}) and
(\ref{eq:fc2}).  The corresponding inflation scheme is
\begin{eqnarray}
S_0=b,\quad S_1=b^{m-k}a,\quad S_{l+1}=S^m_lS_{l-1},
\end{eqnarray}
which is the same as that of $\text{FS}(m,1)$ apart from the different
second initial word.

\subsubsection{The Fibonacci sequence}

Let us commence with the simplest example $\text{FS}(1,1)$.  We
consider the case that the two letters $a$ and $b$ correspond to two
basic unimodular transfer matrices $A$ and $B$, respectively. Denoting
by $A_l$ the total transfer matrix corresponding to a word $S_l$, the
matrix equivalent of Eq.~(\ref{eq:fs}) for FS(1,1) is
\begin{equation}
A_{l+1}=A_{l-1} A_l,
\end{equation}
where $A_1=A$ and $A_0=B$ are the transfer matrices of the two
building blocks $a$ and $b$.  Note the reversed order of matrix
multiplication as compared to the concatenation of letters in
Eq.~(\ref{eq:fs}), which occurs in the related tight-binding model
that is usually considered, whereas the order of matrix
multiplications is not reversed in the optical problem, compare
Eq.~(\ref{eq:opt}).  The well-known trace map reads\cite{Kohmoto83}
\begin{equation}
x_{l+1}=x_{l-1}x_{l}-x_{l-2}.
\label{eq:xfib}
\end{equation}
Note that in part of the literature a factor $1/2$ is introduced in
the definition of $x_l$. Here, we omitted this factor to keep symmetry
between trace and antitrace.  {}From Eq.~(\ref{eq:yab1}), we obtain
the antitrace map
\begin{eqnarray}
y_{l+1}=x_ly_{l-1}+y_{l-2}.
\label{eq:yfib}
\end{eqnarray}
The coefficients of the trace map are constants; however, those of the
antitrace map include the traces. So, if we want to derive the
antitrace map, the trace map must also be known. This is why we have
to consider trace and antitrace maps at the same time.

\subsubsection{Generalized Fibonacci sequences ${\rm FS}(m,n)$}

For FS$(m,n)$, Eq.~(\ref{eq:fs}), the recursion relation for the
transfer matrix is given by
\begin{eqnarray}
A_{l+1}&=&A_{l-1}^n A_l^m\nonumber\\
&=&\left(U_n^{(l-1)}A_{l-1}-U_{n-1}^{(l-1)}I\right)\nonumber\\
&&\times
\left(U_m^{(l-1)}A_{l}-U_{m-1}^{(l-1)}I\right).
\label{eq:aaa}
\end{eqnarray}
Here, we used Eq.~(\ref{eq:an}) and the definition of the functions
$U_{n}(x)=C_{n-1}(x/2)$ given in Appendix~\ref{sec:rum} in terms of
the Chebyshev polynomials of the second kind $C_n(x)$.  Furthermore,
we introduced the notation
\begin{equation}
U_{n}^{(l)} = U_{n}(x_{A_{l}}).
\end{equation}
{}From Eqs.~(\ref{eq:aaa}), (\ref{eq:u}), and (\ref{eq:yab1}), 
the trace and the antitrace maps are obtained as
\begin{eqnarray}
x_{l+1}&=&U_n^{(l-1)}U_m^{(l)}v_l-U_{n-1}^{(l-1)}U_{m+1}^{(l)}
\nonumber\\
&&-U_{n+1}^{(l-1)}U_{m-1}^{(l)},
\label{eq:xl}\\
v_{l+1}&=&U_n^{(l-1)}U_{m+1}^{(l)}v_l-U_{n-1}^{(l-1)}U_{m+2}^{(l)}
\nonumber\\
&&-U_{n+1}^{(l-1)}U_{m}^{(l)},
\label{eq:vl}\\
y_{l+1}&=&U_n^{(l-1)}\left(U_m^{(l)}w_l-U_{m-1}^{(l)}y_{l-1}\right)
\nonumber\\
&&-U_{n-1}^{(l-1)}U_m^{(l)}y_l,
\label{eq:yl}\\       
w_{l+1}&=&U_n^{(l-1)}\left(U_{m-1}^{(l)}w_l-U_{m-2}^{(l)}y_{l-1}\right)
\nonumber\\   
&&+\left(x_{l+1}-U_{n-1}^{(l-1)}U_{m-1}^{(l)}\right)y_l,
\label{eq:wl}
\end{eqnarray}
where $v_l=x_{A_{l-1}A_{l}}$ and $w_l=y_{A_{l-1}A_{l}}$.  Note that the
roles of $v_l$ and $w_l$ are subsidiary.  Eqs.~(\ref{eq:xl}) and
(\ref{eq:vl}) constitute the trace map, Eqs.~(\ref{eq:yl}) and
(\ref{eq:wl}) give the corresponding antitrace map.

For special cases, these expressions simplify considerably.  For
$\text{FS}(1,n)$, we obtain, using the properties (\ref{eq:u}) of the
functions $U_{n}(x)$,
\begin{eqnarray}
x_{l+1}&=&U_n^{(l-1)}v_l-U_{n-1}^{(l-1)}x_l,
\label{eq:xlm1}\\
v_{l+1}&=&U_n^{(l-1)}x_lv_l-U_{n-1}^{(l-1)}\left(x_l^2-1\right)
\nonumber\\
&&-U_{n+1}^{(l-1)},
\label{eq:vlm1}\\
y_{l+1}&=&U_n^{(l-1)}w_l-U_{n-1}^{(l-1)}y_l,
\label{eq:ylm1}\\
w_{l+1}&=&x_{l+1}y_l+U_n^{(l-1)}y_{l-1}.
\label{eq:wlm1}
\end{eqnarray}
Similarly, for $\text{FS}(m,1)$, we find
\begin{eqnarray}
x_{l+1}&=&U_m^{(l)}v_l-U_{m-1}^{(l)}x_{l-1},
\label{eq:xln1}\\
v_{l+1}&=&U_{m+1}^{(l)}v_l-U_{m}^{(l)}x_{l-1},
\label{eq:vln1}\\
y_{l+1}&=&U_m^{(l)}w_l-U_{m-1}^{(l)}y_{l-1},
\label{eq:yln1}\\
w_{l+1}&=&x_{l+1}y_l+U_{m-1}^{(l)}w_l-U_{m-2}^{(l)}y_{l-1}.
\label{eq:wln1}
\end{eqnarray}
Eqs.~(\ref{eq:xln1})--(\ref{eq:wln1}) are quite different from
Eqs.~(\ref{eq:xlm1})--(\ref{eq:wlm1}) above. The corresponding
aperiodic sequences show rather different physical
properties.\cite{Dulea90} We also point out that the trace and
antitrace maps for the sequences $\text{FC}(m,k)$ are given by
Eqs.~(\ref{eq:xln1})--(\ref{eq:wln1}) since they have the same
inflation scheme.

Eliminating the subsidiary variables $v_l$ and $w_l$ in
Eqs.~(\ref{eq:xl})--(\ref{eq:wl}) for the general case
$\text{FS}(m,n)$, we obtain
\begin{eqnarray}
x_{l+1}&=&\frac{U_m^{(l)}U_n^{(l-1)}}{U_m^{(l-1)}}
\left(U_{m+1}^{(l-1)}x_{l}-U_{n+1}^{(l-2)}+U_{n-1}^{(l-2)}\right)
\nonumber\\
&&-U_{m+1}^{(l)}U_{n-1}^{(l-1)}-U_{m-1}^{(l)}U_{n+1}^{(l-1)},
\label{eq:xxl}\\
y_{l+1}&=& \frac{U_{m}^{(l)}U_n^{(l-1)}}{U_m^{(l-1)}}
\left(U_n^{(l-2)}y_{l-2}+U_{m-1}^{(l-1)}y_l\right)
\nonumber\\
&&+U_{m+1}^{(l)}U_n^{(l-1)}y_{l-1}-U_m^{(l)}U_{n-1}^{(l-1)}y_l.
\label{eq:yyl}
\end{eqnarray}
Here, we used Eq.~({\ref{eq:u}}) to simplify the result. The above two
equations are alternative forms of the trace and the antitrace map.

Again, for the special cases $m=1$ or $n=1$, these equations simplify.
For $\text{FS}(1,n)$, we find
\begin{eqnarray}
x_{l+1}&=&{U_n^{(l-1)}}\left(U_{n-1}^{(l-2)}-U_{n+1}^{(l-2)}\right)
+U_{n+1}^{(l-1)}x_l,\nonumber\\
\label{eq:xxlm1}\\
y_{l+1}&=&{U_n^{(l-1)}}\left(U_n^{(l-2)}y_{l-2}+x_ly_{l-1}\right)
-U_{n-1}^{(l-1)}y_l.
\nonumber\\
\label{eq:yylm1}
\end{eqnarray}
The result for $\text{FS}(m,1)$ reads
\begin{eqnarray}
x_{l+1}&=&\frac{U_m^{(l)}}{U_m^{(l-1)}}
\left(U_{m+1}^{(l-1)}x_{l}-x_{l-2}\right)
-U_{m-1}^{(l)}x_{l-1},
\nonumber\\
\label{eq:xxln1}\\
y_{l+1}&=& \frac{U_{m}^{(l)}}{U_m^{(l-1)}}
\left(y_{l-2}+U_{m-1}^{(l-1)}y_l\right)
+U_{m+1}^{(l)}y_{l-1}.
\nonumber\\
\label{eq:yyln1}
\end{eqnarray}
For $\text{FS}(1,1)$, Eqs.~(\ref{eq:xxl})--(\ref{eq:yyln1}) reduce to
Eqs.~(\ref{eq:xfib}) and (\ref{eq:yfib}), as expected. For some other
special values of $m$ and $n$, the trace and antitrace maps of
$\text{FS}(m,n)$ are given in Appendix~\ref{sec:mms}.

\subsection{Generalized Thue-Morse sequences}

Another type of aperiodic sequence is the celebrated TMS and its
generalizations.\cite{Kolar91,GTMS} Here, we consider generalized
sequences $\text{TMS}(m,n)$ with inflation scheme\cite{GTMS}
\begin{equation}
b\rightarrow b^ma^n,\quad a\rightarrow a^nb^m.
\end{equation}
Equivalently, $\text{TMS}(m,n)$  can be constructed as
\begin{eqnarray}
S_0&=&b, \quad \tilde{S}_0=a,\nonumber\\
S_{l+1}&=&S^m_l\tilde{S}^n_{l},\quad 
\tilde{S}_{l+1}=\tilde{S}^n_{l}S^m_l.
\end{eqnarray}
For $m=n=1$, this reduces to the standard TMS.  The recursion relation
for transfer matrices of $\text{TMS}(m,n)$ reads
\begin{equation}
A_{l+1}=B_l^nA_l^m,\quad B_{l+1}=A_l^mB_l^n \label{eq:tmsmn}
\end{equation}
where $A_0$ is the matrix corresponding to the building block $b$, and
$B_0$ corresponds to $a$, respectively.

Using the same method as above, we get
\begin{eqnarray}
x_{l+1}&=&U_n^{(l)}U_m^{(l)}v_l-U_{n-1}^{(l)}U_{m+1}^{(l)}
-U_{n+1}^{(l)}U_{m-1}^{(l)},
\nonumber\\ 
\label{eq:mnx}\\
v_{l+1}&=&U_{2n}^{(l)}U_{2m}^{(l)}v_l-U_{2n-1}^{(l)}U_{2m+1}^{(l)}
-U_{2n+1}^{(l)}U_{2m-1}^{(l)},
\nonumber\\
\label{eq:mnv}
\end{eqnarray}
where $v_l=x_{B_lA_l}$. These two equations determine the trace map.

It is somewhat more complicated to derive the antitrace map because
$y_{A_l}\neq y_{B_l}$.  We define $y_l=y_{A_l}$ and
$\tilde{y}_l=y_{B_l}$. Then, from Eqs.~(\ref{eq:tmsmn}),(\ref{eq:an}),
(\ref{eq:gen10}), and (\ref{eq:gen11}), we have
\begin{eqnarray}
y_{l+1}&=&U_m^{(l)}\left(U_n^{(l)}w_l-U_{n-1}^{(l)}y_l\right)
-U_n^{(l)}U_{m-1}^{(l)}\tilde{y}_l,
\nonumber\\
\label{eq:mny}\\  
\tilde{y}_{l+1}&=&U_m^{(l)}\left(U_n^{(l)}\tilde{w}_l-
U_{n-1}^{(l)}y_l\right)
-U_n^{(l)}U_{m-1}^{(l)}\tilde{y}_l,
\nonumber\\
\\
{w}_{l+1}&=&\left(U_{2n}^{(l)}U_m^{(l)}v_l-U_{2n-1}^{(l)}U_{m+1}^{(l)}
\right.\nonumber\\
&&\left.-U_{2n+1}^{(l)}U_{m-1}^{(l)}\right)U_m^{(l)}{y_l}
+U_{2n}^{(l)}\tilde{y}_l,\\ 
\tilde{w}_{l+1}&=&\left(U_n^{(l)}U_{2m}^{(l)}v_l-U_{n-1}^{(l)}U_{2m+1}^{(l)}
\right.\nonumber\\
&&\left.-U_{n+1}^{(l)}U_{2m-1}^{(l)}\right)U_n^{(l)}\tilde{y}_l+
U_{2m}^{(l)}y_l.
\label{eq:mnw}
\end{eqnarray}
Here, $w_l=y_{B_lA_l}$ and $\tilde{w}_l=y_{A_lB_l}$.  The antitrace map
is completely determined by Eqs.~(\ref{eq:mnx})--(\ref{eq:mnw}).

For $n=1$ and $m=1$, Eqs.~(\ref{eq:mnx})--(\ref{eq:mnw}) reduce to
\begin{eqnarray}
x_{l+1}&=&v_l,\\
v_{l+1}&=&x_l^2(v_l-2)+2,\\
y_{l+1}&=&w_l,\\
\tilde{y}_{l+1}&=&\tilde{w}_l,\\
w_{l+1}&=&x_l[(x_{l+1}-1)y_l+\tilde{y}_l],\\
\tilde{w}_{l+1}&=&x_l[(x_{l+1}-1)\tilde{y}_l+{y}_l].
\end{eqnarray}
This yields the well-known trace map of the TMS
\begin{equation}
x_{l+1}=x_{l-1}^2(x_{l}-2)+2, \label{eq:tmsxxx}
\end{equation}
and the antitrace map
\begin{eqnarray}
y_{l+1}&=&x_{l-1}\left[(x_l-1)y_{l-1}+\tilde{y}_{l-1}\right], 
\label{eq:tmsy}\\
\tilde{y}_{l+1}&=&x_{l-1}\left[(x_l-1)\tilde{y}_{l-1}+y_{l-1}\right].
\label{eq:tmsyt}
\end{eqnarray}
The above two equations give
\begin{eqnarray}
y_{l+1}&=&x_{l-1}[(x_l+x_{l-2}-2)y_{l-1}
\nonumber\\
&&+x_{l-3}x_{l-2}(2-x_{l-2})y_{l-3}]
\end{eqnarray}
which is an alternative form of the antitrace map.

{}From Eqs.~(\ref{eq:mnx})--(\ref{eq:mnw}), we can solve for the
subsidiary quantities $v_l$, $w_{l}$, and $\tilde{w}_{l}$, for
instance,
\begin{eqnarray}
v_{l+1}&=&
\frac{U_{2n}^{(l)}U_{2m}^{(l)}}{U_n^{(l)}U_m^{(l)}}
\left[x_{l} + U_{n-1}^{(l)}\left(U_{m}^{(l)}x_{l}-U_{m-1}^{(l)}\right)
\right.\nonumber\\
&&\left.+ U_{m-1}^{(l)}\left(U_{n}^{(l)}x_{l}-U_{n-1}^{(l)}\right)\right]
\nonumber\\
&&- U_{2n-1}^{(l)}\left(U_{2m}^{(l)}x_{l}-U_{2m-1}^{(l)}\right)
\nonumber\\
&&- U_{2m-1}^{(l)}\left(U_{2n}^{(l)}x_{l}-U_{2n-1}^{(l)}\right).
\label{eq:vvv}
\end{eqnarray}
The combination of Eqs.~(\ref{eq:mnx}) and (\ref{eq:vvv}) gives an
alternative form of the trace map of TMS($m,n$).

\subsection{Period-doubling sequence}

The period-doubling sequence can be generated by the substitution
rule\cite{Luck89}
\begin{eqnarray} 
b&\rightarrow& ba, \quad a\rightarrow b^2,
\end{eqnarray} 
or the inflation scheme 
\begin{eqnarray} 
S_0=b,\quad S_1=ba,\quad S_{l+1}=S_lS^2_{l-1}.
\end{eqnarray}
The inflation scheme is the same as that of
$\text{FS}(1,2)$. Therefore, from Eqs.~(\ref{eq:xxl})--(\ref{eq:yyl}),
the trace and the antitrace maps are obtained as
\begin{eqnarray}
x_{l+1}&=&x_{l-1}(x_lx_{l-1}-x_{l-2}^2+2)-x_l,\\
y_{l+1}&=&x_{l-1}x_{l-2}y_{l-2}+x_lx_{l-1}y_{l-1}-y_l.  
\end{eqnarray}
This yields also the trace and antitrace map for the copper mean
sequence $\text{FS}(1,2)$.

\section{Arbitrary substitution sequences}
\label{sec:ass}

As there exist trace maps for arbitrary substitution
sequences,\cite{Avishai,Iguchi,Kolar902} one natural question is
whether antitrace maps also exist for arbitrary sequences. The answer
is affirmative. We commence our argument in analogy with the
discussion in Ref.~\onlinecite{Avishai} and restrict ourselves to the
case of unimodular matrices.

Let $A_1,A_2,\ldots,A_r$ be $2\!\times\! 2$ matrices and define the
following $2^r$ matrices
\begin{equation}
B_{\epsilon_1\epsilon_2\ldots\epsilon_r}=
A_1^{\epsilon_1}A_2^{\epsilon_2}\ldots A_r^{\epsilon_r}
\end{equation}
where $\epsilon_j\in\{0,1\}$ for $1\le j\le r$.  Then, from
Eq.~(\ref{eq:gen}), any monomial $A_{j_1}A_{j_2}\ldots A_{j_s}$, with
$1\le j_i\le r$ and $1\le i\le s$, can be written as a linear
combination of the matrices
$B_{\epsilon_1\epsilon_2\ldots\epsilon_r}$, namely,\cite{Avishai}
\begin{eqnarray}
\lefteqn{A_{j_1}A_{j_2}\ldots A_{j_s}}
\nonumber\\
&=&\sum_{\epsilon_1=0}^1\sum_{\epsilon_2=0}^1\ldots
\sum_{\epsilon_r=0}^1{c}_{\epsilon_1\epsilon_2\ldots\epsilon_r}
B_{\epsilon_1\epsilon_2\ldots\epsilon_r},
\label{eq:theorem}
\end{eqnarray}
where each coefficient is a polynomial in the traces $x_{A_j}$, $1\le
j\le r$, and the traces $x_{A_jA_k}$, $1\le j<k \le r$.

This result not only yields the trace map, but also gives the
antitrace map for any substitution sequence. We define
\begin{equation}
B_{\epsilon_1\epsilon_2\ldots\epsilon_r,l}=A_{1l}^{\epsilon_1}
A_{2l}^{\epsilon_2}\ldots A_{rl}^{\epsilon_r}
\end{equation}
with $\epsilon_j\in\{0,1\}$, $1\le j\le r$, and $l\ge 0$, where
$A_{jl}$ is the unimodular $2\!\times\! 2$ matrix associated to the
$l$-th iterate of the $j$-th letter. Since each matrix in
$B_{\epsilon_1\epsilon_2\ldots\epsilon_r,l+1}$ is, by definition, a
monomial in the matrices $A_{jl}$, they can be expanded in terms of
the matrices $B_{\epsilon_1\epsilon_2\ldots\epsilon_r,l}$ according to
(\ref{eq:theorem}). Then the trace of
$B_{\epsilon_1\epsilon_2\ldots\epsilon_r,l+1}$ is a polynomial in the
$2^r-1$ traces of $B_{\epsilon_1\epsilon_2\ldots\epsilon_r,l}$; and
the antitrace of $B_{\epsilon_1\epsilon_2\ldots\epsilon_r,l+1}$ is a
polynomial in the $2^r-1$ antitraces of
$B_{\epsilon_1\epsilon_2\ldots\epsilon_r,l}$.  Therefore, we conclude
that both the trace and antitrace maps exist for arbitrary
substitution sequences, and the dimension of the antitrace map is
$2^r-1$. Next, we present a concrete example to illustrate this
conclusion.

The Rudin-Shapiro sequence can be defined by means of a substitution
rule on four letters.\cite{RS} The substitution rule is
\begin{equation}
a\rightarrow ac, \quad b\rightarrow dc,\quad c\rightarrow ab,
\quad d\rightarrow db,
\end{equation}
and the corresponding matrix recursion relations are
\begin{eqnarray}
A_{l+1}&=&C_{l}A_{l}, \quad 
B_{l+1}=C_{l}D_{l}, \nonumber\\
C_{l+1}&=&B_{l}A_{l}, \quad
D_{l+1}=B_{l}D_{l}. 
\end{eqnarray}
We have the useful relation\cite{Avishai}
\begin{equation}
D_l=C_{l}A_l^{-1}B_l,
\end{equation}
which effectively reduces the sequence to three basic letters.  Now,
we choose the seven matrices $A_l$, $B_l$, $C_l$, $D_l$, $A_lC_l$,
$A_lB_l$, and $B_lC_l$ as our basic set of matrices.

In what follows, we denote the traces and antitraces by
\begin{eqnarray}
\lefteqn{
a_{l}=x_{A_{l}},\quad
b_{l}=x_{B_{l}},\quad
c_{l}=x_{C_{l}},\quad
d_{l}=x_{D_{l}},}\nonumber\\
\lefteqn{
e_{l}=x_{A_{l}C_{l}},\quad 
f_{l}=x_{A_{l}B_{l}},\quad 
g_{l}=x_{B_{l}C_{l}},}\nonumber\\
\lefteqn{
\tilde{a}_{l}=y_{A_{l}},\quad
\tilde{b}_{l}=y_{B_{l}},\quad
\tilde{c}_{l}=y_{C_{l}},\quad
\tilde{d}_{l}=y_{D_{l}},}\nonumber\\
\lefteqn{
\tilde{e}_{l}=y_{A_{l}C_{l}},\quad 
\tilde{f}_{l}=y_{A_{l}B_{l}},\quad 
\tilde{g}_{l}=y_{B_{l}C_{l}}.}
\end{eqnarray}
By using Eqs.~(\ref{eq:gen}) and (\ref{eq:gen10}), we obtain
\begin{eqnarray}
A_{l+1}&=&C_{l}A_{l},
\nonumber\\
B_{l+1}&=&c_{l}D_{l}-a_{l}B_{l}+A_{l}B_{l},
\nonumber\\
{C}_{l+1}&=&B_{l}A_{l},
\nonumber\\
{D}_{l+1}&=&(a_{l}g_{l}-c_{l}f_{l})B_{l}-c_{l}A_{l}+b_{l}{D_{l}}
\nonumber\\
&&-(g_{l}-b_{l}c_{l}){A_{l}B_{l}}
\nonumber\\
&&+(f_{l}-a_{l}b_{l}){C_{l}B_{l}}+{C_{l}A_{l}},
\nonumber\\
A_{l+1}{C}_{l+1}&=&f_{l}{C_{l}A_{l}}-b_{l}{C_{l}}+{C_{l}B_{l}},
\nonumber\\
A_{l+1}B_{l+1}&=&c_{l}[(1-a_{l}^{2})B_{l}+e_{l}{D_{l}}+a_{l}{A_{l}B_{l}}]
\nonumber\\
&&-{C_{l}B_{l}},
\nonumber\\
B_{l+1}{C}_{l+1}&=&b_{l}c_{l}[(f_{l}a_{l}-a_{l}^{2}b_{l}+b_{l}){C_{l}}
\nonumber\\
&&+a_{l}{D_{l}}-{C_{l}B_{l}}-(f_{l}-a_{l}b_{l}){C_{l}A_{l}}]
\nonumber\\
&&-b_{l}[(f_{l}-a_{l}b_{l})(a_{l}I-A_{l})
\nonumber\\
&&+b_{l}I+(a_{l}^{2}-1)B_{l}-a_{l}{A_{l}B_{l}}]
\nonumber\\
&&-c_{l}{C_{l}}+{I}.
\label{eq:mat}
\end{eqnarray}
Note that the order of multiplication of two matrices on the
right-hand sides of these equations may differ from the order of our
basic matrix products $A_lC_l$, $A_lB_l$, or $B_lC_l$. We can use
Eq.~(\ref{eq:gen}) to reverse the order to obtain a systems of
equations that closes with our seven basic matrices.

Now, from Eq.~(\ref{eq:mat}) and (\ref{eq:yab1}), the trace and
antitrace maps are obtained as
\begin{eqnarray}
a_{l+1}&=&e_{l},
\nonumber\\
b_{l+1}&=&c_{l}d_{l}-a_{l}b_{l}+f_{l},
\nonumber\\
c_{l+1}&=&f_{l},
\nonumber\\
d_{l+1}&=&b_{l}d_{l}-a_{l}c_{l}+e_{l},
\nonumber\\
e_{l+1}&=&e_{l}f_{l}-b_{l}c_{l}+g_{l},
\nonumber\\
f_{l+1}&=&c_{l}(d_{l}e_{l}-a_{l}^{2}b_{l}+a_{l}f_{l}+b_{l})-g_{l},
\nonumber\\
g_{l+1}&=&b_{l}c_{l}(a_{l}c_{l}f_{l}-e_{l}f_{l}-a_{l}^{2}b_{l}c_{l}
          +a_{l}b_{l}e_{l}+b_{l}c_{l}
\nonumber\\
&&+a_{l}d_{l}+g_{l})-b_{l}^{2}-c_{l}^{2}+2,
\nonumber\\
\tilde{a}_{l+1}&=&c_{l}\tilde{a}_{l}+a_{l}\tilde{c}_{l}-\tilde{e}_{l},
\nonumber\\
\tilde{b}_{l+1}&=& c_{l}\tilde{d}_{l}-a_{l}\tilde{b}_{l}+\tilde{f}_{l},
\nonumber\\
\tilde{c}_{l+1}&=&b_{l}\tilde{a}_{l}+a_{l}\tilde{b}_{l}-\tilde{f}_{l},
\nonumber\\
\tilde{d}_{l+1}&=&(a_{l}g_{l}-a_{l}b_{l}c_{l})\tilde{b}_{l}+b_{l}\tilde{d}_{l}-
                  (g_{l}-b_{l}c_{l})\tilde{f}_{l}
\nonumber\\
&&+(f_{l}-a_{l}b_{l})(b_{l}\tilde{c}_{l}-\tilde{g}_{l})
              +a_{l}\tilde{c}_{l}-\tilde{e}_{l},
\nonumber \\
\tilde{e}_{l+1}&=&f_{l}(c_{l}\tilde{a}_{l}+a_{l}\tilde{c}_{l}-\tilde{e}_{l})
                  +c_{l}\tilde{b}_{l}-\tilde{g}_{l},
\nonumber\\
\tilde{f}_{l+1}&=&c_{l}(-a_{l}^2\tilde{b}_{l}+e\tilde{d}_{l}
                  +a_{l}\tilde{f}_{l})-b_{l}\tilde{c}_{l}+\tilde{g}_{l},
\nonumber\\
\tilde{g}_{l+1}&=&b_{l}(1-c_{l}^{2})(f_{l}-a_{l}b_{l})\tilde{a}_{l} +
                  b_{l}(1-a_{l}^{2}-c_{l}^{2})\tilde{b}_{l}
\nonumber\\
               &&-c_{l}\tilde{c}_{l}+a_{l}b_{l}c_{l}\tilde{d}_{l}
                 +b_{l}c_{l}(f_{l}-a_{l}b_{l})\tilde{e}_{l} 
                 +a_{l}b_{l}\tilde{f}_{l}
\nonumber\\
               &&+b_{l}c_{l}\tilde{g}_{l}.
\label{eq:below}
\end{eqnarray}
Thus, we derived the trace and antitrace maps of the Rudin-Shapiro
sequence.

Now, we discuss the dimension of the antitrace map. Let $A$, $B$, and
$C$ be $2\times 2$ matrices. Then\cite{Avishai}
\begin{eqnarray}
ABC\!
&=&\left[(x_{ABC}-x_{AB}x_{C}-x_Ax_{BC}+x_Ax_Bx_{C}){I}\right.
\nonumber\\
&&+(x_{BC}-x_Bx_{C})A-x_{AC}B
\nonumber\\
&&+(x_{AB}-x_Ax_B){C}
\nonumber\\
&&\left.+x_{C}{AB}+x_B{AC}+x_A{BC}\right]/2.
\label{eq:abc}
\end{eqnarray} 
Taking the trace on both sides of Eq.~(\ref{eq:abc}), we are led to a
trivial identity. However, if we take the antitrace, we obtain
\begin{eqnarray}
y_{ABC}&=&\left[(x_{BC}-x_Bx_{C})y_A-x_{AC}y_B\right.
\nonumber\\
&&+(x_{AB}-x_Ax_B)y_{C}
\nonumber\\
&&\left.+x_{C}y_{AB}+x_By_{AC}+x_Ay_{BC}\right]/2.
\label{eq:yabc}
\end{eqnarray}
The antitrace of any monomial can be written as a linear combination
of a polynomial in the antitraces $y_{A_j}$, $1\le j\le r$, and the
antitraces $y_{A_jA_k}$, $1\le j<k \le r$. Each coefficient is a
polynomial in the traces $x_{A_j}$, $1\le j\le r$ and the traces
$x_{A_jA_k}$, $1\le j<k \le r$.  {}From this observation we conclude
that the dimension of our antitrace map is $r(1+r)/2$, {\it i.e.}, the
dimension is reduced from $2^r-1$. Here, for the dimension of the
antitrace map, we do not take into account the dimension of the trace
map, which enters the coefficients of the antitrace map. Thus, the
full dimension of the trace and antitrace map is given by the sum of
their respective dimensions.

Let us consider two ternary sequences as examples.\cite{YYLiu91,Ali88}
Our first example of a three-letter substitution rule and the
corresponding recursion relation for the transfer matrices
is\cite{YYLiu91}
\begin{eqnarray}
\lefteqn{a\rightarrow b, \quad b\rightarrow c,\quad c\rightarrow ca,}
\nonumber\\
\lefteqn{A_{l+1}=B_l,\quad B_{l+1}=C_l,\quad C_{l+1}=A_lC_l.}
\end{eqnarray}
Using Eqs.~(\ref{eq:a2}) and (\ref{eq:gen10}), we obtain
\begin{eqnarray}
A_{l+1}&=&B_{l},
\nonumber\\
B_{l+1}&=&C_{l},
\nonumber\\
C_{l+1}&=&A_{l}C_{l},
\nonumber\\
B_{l+1}A_{l+1}&=&C_{l}B_{l},
\nonumber\\
C_{l+1}B_{l+1}&=&x_{C_l}A_{l}C_{l}-A_{l},
\nonumber\\
A_{l+1}C_{l+1}&=&B_{l}A_{l}C_{l},
\nonumber\\
B_{l+1}A_{l+1}C_{l+1}&=&B_{l}A_{l}+x_{B_lA_lC_l}C_{l}
\nonumber\\
                     & & -x_{B_lA_l}I.
\label{eq:ter1}
\end{eqnarray}
Taking the trace of the above equation, we obtain the trace map. The
dimension of the trace map is $2^3-1=7$.  Taking the antitrace of the
sixth line of the above equation, we can expand it according to
Eq.~(\ref{eq:yabc}).  Therefore, the last equality in
Eq.~(\ref{eq:ter1}) is not necessary, and the dimension of the
antitrace map is $3(3+1)/2=6$.

Our second example is the three-component FS generated by\cite{Ali88}
\begin{eqnarray}
\lefteqn{a\rightarrow b,\quad b\rightarrow c,\quad
c\rightarrow abc,}\nonumber\\
\lefteqn{A_{l+1}=B_l,\quad B_{l+1}=C_l,\quad C_{l+1}=C_lB_lA_l.}   
\end{eqnarray}
The corresponding maps for the matrices are
\begin{eqnarray}
A_{l+1}&=&B_{l},
\nonumber\\
B_{l+1}&=&C_{l},
\nonumber\\
C_{l+1}&=&C_{l}B_{l}A_{l},
\nonumber\\
B_{l+1}A_{l+1}&=&C_{l}B_{l},
\nonumber\\
C_{l+1}B_{l+1}&=&B_{l}A_{l}-x_{B_lA_l}I+x_{C_lB_lA_l}C_{l},
\nonumber\\
C_{l+1}B_{l+1}A_{l+1}&=&A_{l}-x_{A_l}I+x_{C_lB_lA_l}C_{l}B_{l}.
\label{eq:ter11}
\end{eqnarray}
We see that, for this particular sequence, both the trace and
antitrace maps are six-dimensional.

\section{Maps for matrix elements}
\label{sec:mem}

As discussed in Sec.~\ref{sec:intro}, we need to know all elements of
the global transfer matrix in order to compute certain physical
quantities. Thus, the trace and antitrace maps may not be sufficient,
and one would like to determine analogous maps for each of the matrix
elements.

Actually, from Eq.~(\ref{eq:theorem}), we know that such matrix
element maps exist for any substitution rule, and Eqs.~(\ref{eq:mat}),
(\ref{eq:ter1}), and (\ref{eq:ter11}) already contain examples of
matrix element maps. Now, we investigate the maps for the matrix
elements of the FS($m$,$n$) and TMS($m$,$n$).

Using Eqs.~(\ref{eq:gen}), (\ref{eq:gen10}), and (\ref{eq:gen11}), we
obtain the matrix maps of FS($m,n$) as (\ref{eq:aaa}) and
\begin{eqnarray}
{M}_{l+1}&=&U_n^{(l-1)}U_{m-1}^{(l)}{M}_l\nonumber\\
&&+\left(x_{l+1}-U_{n-1}^{(l-1)}U_{m-1}^{(l)}\right)A_l\nonumber\\
&&-U_n^{(l-1)}U_{m-2}^{(l)}A_{l-1}\nonumber\\
&&+\left(U_{n-1}^{(l-1)}U_{m-2}^{(l)}+v_{l+1}-x_lx_{l+1}\right){I},
\label{eq:matmap1} 
\end{eqnarray}
where $M_l=A_{l-1}A_l$. The traces $x_{l+1}$ and $v_{l+1}$ appearing on
the right-hand side of Eq.~(\ref{eq:matmap1}) are given in terms of
$v_l$ via Eqs.~(\ref{eq:xl}) and (\ref{eq:vl}), respectively.

Similarly, the matrix map of TMS($m$,$n$) is obtained as
\begin{eqnarray}
{A_{l+1}}&=&U_n^{(l)}\left(U_m^{(l)}{N}_l-U_{m-1}^{(l)}B_l\right)
\nonumber\\
&&               -U_{n-1}^{(l)}\left(U_{m}^{(l)}A_l-U_{m-1}^{(l)}I\right),
\label{eq:matmap20}\\
{B_{l+1}}&=&U_n^{(l)}\left(U_m^{(l)}{\tilde{N}}_l-U_{m-1}^{(l)}B_l\right)
\nonumber\\
&&          -U_{n-1}^{(l)}\left(U_{m}^{(l)}A_l-U_{m-1}^{(l)}I\right),
\\
N_{l+1}&=&\left(U_{2n}^{(l)}U_{m}^{(l)}v_l-U_{2n-1}^{(l)}U_{m+1}^{(l)}
\right.\nonumber\\
&&\left.-U_{2n+1}^{(l)}U_{m-1}^{(l)}\right)
\left(U_m^{(l)}A_l-U_{m-1}^{(l)}I\right)
\nonumber\\
&&+U_{2n}^{(l)}B_l-U_{2n+1}^{(l)}I,
\\
\tilde{N}_{l+1}&=&
\left(U_n^{(l)}U_{2m}^{(l)}v_l-U_{n-1}^{(l)}U_{2m+1}^{(l)}\right.
\nonumber\\
&&\left. -U_{n+1}^{(l)}U_{2m-1}^{(l)}\right)
   \left(U_n^{(l)}B_l-U_{n-1}^{(l)}I\right)
\nonumber\\
&&+U_{2m}^{(l)}A_l-U_{2m+1}^{(l)}I, 
\label{eq:matmap2}
\end{eqnarray}
where ${N}_l=B_lA_l$ and ${\tilde{N}}_l=A_lB_l$. We can eliminate the
subsidiary matrices $M_l$, $N_l$, and $\tilde{N}_l$ from
Eqs.~(\ref{eq:aaa}) and (\ref{eq:matmap1})--(\ref{eq:matmap2}). For
example, Eq.~(\ref{eq:matmap1}) becomes
\begin{eqnarray}
M_l&=&\frac{1}{U_m^{(l-1)}}\left(U_{m-1}^{(l-1)}A_l
+U_n^{(l-2)}A_{l-2}-U_{n-1}^{(l-2)}I\right)
\nonumber\\
&&+v_lI-x_lx_{l-1}I+x_lA_{l-1}.
\label{eq:smat}
\end{eqnarray} 
Thus, we obtain another form of the matrix map of FS($m,n$) given by
Eqs.~(\ref{eq:aaa}) and (\ref{eq:smat}).

For $m=n=1$, Eqs.~(\ref{eq:aaa}) and (\ref{eq:smat}) reduce to
\begin{eqnarray}
A_{l+1}=(x_{l+1}-x_lx_{l-1}){ I}+x_lA_{l-1}+A_{l-2}.
\label{eq:matmap3}
\end{eqnarray}
This is the matrix map of the FS. For the TMS, we find from
Eqs.~(\ref{eq:matmap20})--(\ref{eq:matmap2}) for $m=n=1$
\begin{eqnarray} 
A_{l+1}&=& x_{l-1}[(x_l-1){A}_{l-1}
+B_{l-1}-x_{l-1}I]+I,
\nonumber\\
\label{eq:matmap40}\\ 
B_{l+1}&=& x_{l-1}[(x_l-1)B_{l-1}
+A_{l-1}-x_{l-1}I]+I.
\nonumber\\
\label{eq:matmap4}
\end{eqnarray} 
The maps for the matrix elements are easily obtained from the matrix
map, thus we do not give them explicitly.

Specifically, we consider the FS. {}From Eq.~(\ref{eq:matmap3}), it is
interesting to find that the maps for the non-diagonal elements, and
for the difference of the diagonal elements, coincide with the
antitrace map, Eq.~(\ref{eq:yfib}).  {}From Eqs.~(\ref{eq:matmap40})
and (\ref{eq:matmap4}), this fact also holds for the TMS. Actually, as
again follows from Eq.~(\ref{eq:theorem}), we arrive at the important
conclusion that the maps for the antitrace, the non-diagonal elements,
and the difference of the diagonal elements are all the same for
arbitrary substitution rules. This means that the knowledge of the
trace and antitrace maps suffices to compute any physical quantities
related to the global transfer matrix.

\section{Applications}
\label{sec:app}

We now turn our attention to applications of the dynamical map method
developed in this paper. In what follows, we are going to consider three
examples.

\subsection{Optical Multilayers}

As our first example, we show how to use the antitrace map to
calculate light transmission coefficients.

The transmission of light through aperiodic multilayers arranged
according to the FS,\cite{Optics,Jin96} the ``non-Fibonacci''
sequence,\cite{Dulea90,Riklund88} the TMS,\cite{Liu97} and the
generalized TMSs\cite{Kolar91} was studied in the literature.
Possible applications of quasiperiodic multilayers as optical switches
and memories have been suggested by Schwartz.\cite{Schwartz88} Huang
{\it et al}\/\cite{Huang93} and Yang {\it et al}\/\cite {XBYang99}
found an interesting switch-like property in the light transmission
through a FC($m$) multilayer.

Using the antitrace map, we re-investigate the light transmission
through FC($m$) which is sandwiched by two media of type $b$. In
analogy with the discussion of Ref.~\onlinecite{XBYang99}, we write
the corresponding transfer matrices as
\begin{eqnarray}
A_1&=&{P}_b,
\nonumber\\
A_2&=&{P}_{b}^{m-1}{P}_{ba}{P}_a{P}_{ab},
\nonumber\\
A_{l+1}&=&A_l^mA_{l-1}.
\label{eq:fcm}
\end{eqnarray}
The recursion relation for the transfer matrix (\ref{eq:fcm}) is a
little different from Eq.~(\ref{eq:fs}) for FS($m$,1).  It can easily
be seen that the trace map is the same, but that the antitrace map
differs slightly. The antitrace map is given by
\begin{eqnarray}
y_{l+1}&=&U_m^{(l)}\bar{w}_l-U_{m-1}^{(l)}y_{l-1},\\
\bar{w}_{l+1}&=&x_{l+1}y_l+U_{m-1}^{(l)}\bar{w}_l-U_{m-2}^{(l)}y_{l-1}
\end{eqnarray}
where $\bar{w}=y_{A_{l}A_{l-1}}$.

We consider the case that the light vertically transmits the
multilayer and choose the thicknesses of the layers $d_a$ and $d_b$
appropriately in order to make $n_ad_a=n_bd_b$. Then, we have phase
differences $\delta_a=\delta_b=\delta$, compare Eq.~(\ref{eq:phase}).
For $\delta=(n+1/2)\pi$, the propagation matrices become
\begin{eqnarray}
{P}_{a}&=&{P}_{b}=\left(\matrix {0&-1\cr 1&0} \right).
\end{eqnarray}
{
}From the above equation and Eqs.~(\ref{eq:fcm}), (\ref{eq:an}), and
(\ref{eq:u}), we can obtain the initial conditions for the trace and
antitrace maps as
\begin{eqnarray}
x_1&=&0,
\nonumber\\
x_2&=&-U_{m-1}(0)\varrho_{1},
\nonumber\\
v_2&=&U_{m-2}(0)\varrho_{1},
\nonumber\\
y_1&=&2,
\nonumber\\
y_2&=&-U_{m-2}(0)\varrho_{1},
\nonumber\\
\bar{w}_2&=&-U_{m-1}(0)\varrho_{1},
\end{eqnarray}
where
\begin{equation}
\varrho_{m}=R^{m}+R^{-m},\quad R=n_{a}/n_{b}.
\label{eq:rho}
\end{equation}
The initial conditions depend on the parameters $\varrho_{1}$ and $m$,
while the recursion relations only depend on $m$. {}From
Eq.~(\ref{eq:u}), we know that
\begin{eqnarray}
U_m(0)&=&\frac{1}{2}[1-(-1)^m]i^{m-1}
\nonumber\\
&=&\cases{0 & for $m=2k$\cr (-1)^k & for $m=2k+1$}.
\end{eqnarray}
Since $U_{m+4}(0)=U_{m}(0)$, {\it i.e.}, the function $U_m(0)$ is
periodic in $m$ with period four, the initial conditions, given in
Table~\ref{tab:1}, also show this periodicity.  The initial conditions
for FC($2q$), $q=1,2,3,\ldots$, or for FC($2q+1$), only differ by the
sign of the parameter $R$.  Thus, it is natural to divide the FC($m$)
into two classes, FC($2q$) and FC($2q+1$).

{}From the initial conditions and recursion equations, we can directly
obtain the trace, the antitrace and the transmission coefficients of
FC($2q$), which are given in Table~\ref{tab:2}. It can be seen that
the trace and the antitrace vanish alternately.  The trace shows
periodicity with period four for odd values of $q$, and period two for
even $q$, but the antitrace shows no periodicity. Thus, the
transmission coefficient also is not periodic in $l$.  For even $l$,
the transmission coefficient does not depend on $m$. However, for odd
$l$, the transmission coefficient depends on $m$ and $l$, see
Table~\ref{tab:2}.

Table~\ref{tab:3} shows the results for FC($2q+1$). In this case, the
trace, the antitrace and the transmission coefficient are periodic in
$l$ with period six.  The transmission coefficients are the same for
$l=2$, $l=3$, and $l=6$ and do not depend on $m$. We find that the
multilayer is transparent for $l=6i+1$, $i=0,1,2...$.

Here, we not only recover the recent results of
Ref.~\onlinecite{XBYang99}, but also give a natural classification of
FC($m$) and derive the periodicities of the trace and antitrace maps.

\subsection{Harmonic Chains}

As our second example, we show how to apply the map for the matrix
elements to calculate some physical quantities for a harmonically
coupled Fibonacci chain. The transmission coefficient and the Lyapunov
exponent were already given in Eqs.~(\ref{eq:ttt}) and
(\ref{eq:Gamma}). We know the trace (\ref{eq:xfib}) and antitrace maps
(\ref{eq:yfib}) for this system. In order to determine the
transmission coefficient, we additionally need to know the map for the
difference $z_l$ of the diagonal elements in Eq.~(\ref{eq:ttt}). As
discussed in Sec.~\ref{sec:mem}, the map for $z_l$ is the same as the
map for the antitrace $y_l$.

Now, this leaves us with the problem to determine the initial
conditions. By a so-called transfer matrix
``renormalization'',\cite{Macia99} the transfer matrix product can be
re-written in terms of ``renormalized'' transfer matrices such that
these are arranged according to the FS.  Following the discussion in
Ref.~\onlinecite{Macia99}, we choose a special value of parameters
\begin{equation}
\Omega=\frac{\alpha-2\beta+1}{\alpha(1-\beta)} = 
\frac{m_a\omega^2}{K_{ab}}
\end{equation}
where $\alpha=m_b/m_a$ and $\beta=K_{aa}/K_{ab}$. The first two
renormalized transfer matrices are\cite{Macia99}
\begin{eqnarray}
A_1&=&\left(\matrix{1&0\cr \eta_1&1}\right),
\nonumber\\
A_2&=&\left(\matrix{-1&0\cr \eta_2&-1}\right),
\end{eqnarray}
where $\eta_1=2(\alpha-2)$ and $\eta_2=2(1-\alpha)$.  Note that these
two matrices commute with each other for arbitrary values of $\eta_1$
and $\eta_2$. {}From this equation, we obtain
\begin{equation}
A_3=A_1A_2=\left(\matrix{-1&0\cr \eta_2-\eta_1&-1}\right).
\end{equation}
Thus, the initial conditions are given by
\begin{eqnarray}
\lefteqn{x_1=2,\quad x_2=-2, \quad  x_3=-2,} 
\nonumber\\
\lefteqn{y_1=\eta_1, \quad y_2=\eta_2, \quad y_3=\eta_2-\eta_1,}
\nonumber\\
\lefteqn{z_1=z_2=z_3=0.}
\end{eqnarray}

{}From the antitrace map (\ref{eq:yfib}), we find that $z_l=0$ for all
$l$. Using the trace map (\ref{eq:xfib}), we easily obtain
$x_{3i+1}=2$, $x_{3i+2}=-2$, and $x_{3i}=-2$. That is, the trace map
is periodic in $l$ with period three.  Then, from Eqs.~(\ref{eq:ttt})
and (\ref{eq:Gamma}) the transmission coefficient and the Lyapunov
exponent have the simple forms
\begin{eqnarray}
t_l^{-1}&=&1+\frac{y_l^2}{4\sin^2k},\\
\Gamma_l&=&N^{-1}\ln(y_l^2+2).
\end{eqnarray}
{}From the initial conditions for $y_l$ and the antitrace map
(\ref{eq:yfib}), we easily find that the modulus of $y_l$ is
\begin{equation}
|y_l|=|F_{l}\eta_2-F_{l-1}\eta_1|, \quad l\ge 3,
\end{equation}
where $F_l$ denotes the Fibonacci number defined by the recursion
$F_l=F_{l-1}+F_{l-2}$ with $F_0=F_1=1$.

Finally, the transmission coefficient and the Lyapunov exponent are
obtained as
\begin{eqnarray}
t_l^{-1}&=&
1+\frac{(F_{l}\eta_2-F_{l-1}\eta_1)^2}{4\sin^2k},\\
\Gamma_l&=&N^{-1}\ln[(F_{l}\eta_2-F_{l-1}\eta_1)^2+2].
\end{eqnarray}
Thus, using our matrix element maps, we have re-derived the result of
Ref.~\onlinecite{Macia99}.

\subsection{Electronic systems}

We now apply the trace and antitrace method to the transmission problem 
in electronic systems for the examples of the FS and the TMS. 
In what follows, we choose the parameters
as $\epsilon_a=-\epsilon_b=\epsilon$, $t_{ab}=1$, and $t_{aa}=t_{bb}=t$.

\subsubsection{Fibonacci sequence}

For the FS, there are actually four different local transfer matrices
$M_n$ (\ref{eq:Schroedinger}), because the hopping matrix elements
depend on three subsequent letters in the FS.  Nevertheless, the
transfer matrix product can be re-written\cite{Macia} in terms of two
matrices
\begin{eqnarray}
M_b&=&\left(\matrix {E-\epsilon&
-1 \cr 1&0}\right)\left(\matrix {E+\epsilon&
-1 \cr 1&0}\right), \nonumber\\ 
M_a&=&\left(\matrix {E-\epsilon&
-t \cr 1&0}\right)\left(\matrix {(E-\epsilon)/t&
-1/t \cr 1&0}\right)\nonumber\\
&&\times\left(\matrix {E+\epsilon&
-1 \cr 1&0}\right),
\label{electronic}
\end{eqnarray}
such that the resulting transfer matrix product is again arranged
according to the Fibonacci sequence.

For the trace and antitrace maps, we only need to know the first three
matrices $A_{1}=M_a$, $A_{2}=M_bM_a$, and $A_{3}=M_aM_bM_a$. {}From
Eq.~(\ref{electronic}), these matrices and thus the initial conditions
are easily obtained. In order to obtain an analytical result, we
restrict ourselves to the case $E=\epsilon=0$. For this particular choice
of parameters, Eq.~(\ref{eq:tttttt}) simplifies to
\begin{equation}
t_l=\frac{4}{x_l^2+y_l^2}, \label{eq:tttttt1}
\end{equation}
which is formally the same as Eq.~(\ref{eq:t}). The initial conditions become
\begin{eqnarray}
\lefteqn{x_1=0,\quad x_2=0, \quad  x_3=2,} 
\nonumber\\
\lefteqn{y_1=-t-1/t, \quad y_2=t+1/t, \quad y_3=0.}
\end{eqnarray}
{}From the trace and antitrace map equations
(\ref{eq:xfib})--(\ref{eq:yfib}) for the FS, we can easily find that
both the trace $x_l$ and the antitrace $y_l$ are periodic in $l$ with
period six. In one period the traces are $0$, $0$, $2$, $0$, $0$,
$-2$, and the antitraces are $-t-1/t$, $t+1/t$, $0$, $t+1/t$, $t+1/t$,
$0$. {}From Eq.~(\ref{eq:tttttt1}), we deduce that the transmission
coefficient $t_l$ is periodic in $l$ with period three. For one
period, the transmission coefficients are given by $4/(t+1/t)^2$,
$4/(t+1/t)^2$, and $1$. If the hopping parameter $t=1$, the
transmission coefficient $t_l=1$ for all values of $l$, which is the
trivial (periodic) case. Next we consider the electronic transmission
for the TMS.

\subsubsection{Thue-Morse sequence}

We consider the on-site model for the TMS, i.e., the hopping parameter
$t=1$. So there are only two kinds of transfer matrices
\begin{equation}
B_0=\left(\matrix {E+\epsilon &-1 \cr 1&0}\right), \; 
A_0=\left(\matrix {E-\epsilon &-1 \cr 1&0}\right).
\end{equation}
{}From these, we can calculate the matrices $A_1$, $B_1$, $A_2$, and
$B_2$, and thus the initial conditions for the trace and antitrace
map. Again, in order to obtain an analytical result, we limit
ourselves to the case where the parameter $\epsilon$ and the energy
$E$ fulfill a particular relation, $E=\sqrt{2+\epsilon^2}$. In this case, 
the initial conditions become
\begin{eqnarray}
\lefteqn{x_0=\sqrt{2+\epsilon^2}-\epsilon,\;\;
x_1=0,\;\; x_2=-2-4\epsilon^2,} 
\nonumber\\
\lefteqn{y_0=\tilde{y}_0=2,\;\;
y_1=\tilde{y}_1=2\sqrt{2+\epsilon^2}, \;\;
y_2=-\tilde{y}_2=4\epsilon,}
\nonumber\\
\lefteqn{z_0=\sqrt{2+\epsilon^2}-\epsilon, \;\; 
\tilde{z}_0=\sqrt{2+\epsilon^2}+\epsilon,}
\nonumber\\
\lefteqn{z_1=\tilde{z}_1=2, \;\; 
z_2=-\tilde{z}_2=4\epsilon\sqrt{2+\epsilon^2},}
\end{eqnarray}
where ${z}_l={(A_l)}_{11}-{(A_l)}_{22}$ and
$\tilde{z}_l={(B_l)}_{11}-{(B_l)}_{22}$.  {}From Eq.~(\ref{eq:tmsxxx}),
we deduce that the traces $x_l=2$ for all $l\ge 3$. {}From the
antitrace map equations (\ref{eq:tmsy})--(\ref{eq:tmsyt}) and the
above initial conditions, we easily find that $y_l=z_l=0$ for $l\ge
3$. Thus, we obtain the result that the transmission coefficient
$t_l=1$ for $l\ge 3$. For $l=1$ and $l=2$, the transmission
coefficients are given by $t_1=(2-\epsilon^2)/(2+\epsilon^2)$ and
$t_2=(2-\epsilon^2)/(2+7\epsilon^2+4\epsilon^4)$, respectively.

The examples considered here show that trace and antitrace maps
provide a convenient tool for the computation of physical quantities
related to the global transfer matrices of aperiodic substitution
systems. In the applications presented above, we mainly concentrated
on obtaining analytical results, and therefore had to restrict the
discussion to specific values of the parameters. The trace and
antitrace map equations, of course, are not restricted to these cases,
but there will be no simple closed-form solutions to the recursion
relations in general. The particular parameter values considered above
correspond to periodic orbits of the associated dynamical
systems. These cases, and probably all examples where simple solutions
exist, share the property that, at a certain stage, different transfer
matrices commute with each other, and thus are simultaneously
diagonalizable. This also explains why these systems turn out to be
transparent, because it does not matter in which order one multiplies
matrices that commute with each other. In spite of these comments, the
method presented here is expedient and useful for the investigation of
physical systems built on aperiodic substitution sequences, because
the trace and antitrace map equations can very efficiently be used in
numerical investigations of large, but finite, systems.

\section{Conclusions}
\label{sec:con}

In conclusion, we have extended the well-studied trace-map method for
the investigation of aperiodic substitution systems by considering
corresponding maps for the antitrace and the matrix elements of the
transfer matrices. Our main results are the following.

Firstly, we obtained the trace and antitrace maps for various
aperiodic sequences, such as generalized FSs and TMSs, the
periodic-doubling sequence, examples of ternary sequences, and the
four-letter Rudin-Shapiro sequence. The dimension of the dynamical
systems defined by the trace map and our antitrace maps is $r(r+1)/2$
plus the dimension of the trace map itself, where $r$ denotes the
number of basic letters in the aperiodic sequence.  Secondly, we
showed that trace and antitrace maps can be constructed for arbitrary
substitution rules.  Thirdly, we introduced analogous maps for
specific matrix elements of the transfer matrix, but it turns out that
the maps for the off-diagonal elements and those for the difference of
the diagonal elements coincide with the antitrace map. Thus, from the
trace and antitrace map, we can determine any physical quantity
related to the global transfer matrix of the system. Finally, as examples
of applications of the trace and antitrace map method, we
investigated the transmission problem for optical multilayers,
harmonic chains, and electronic systems arranged according to the FS 
or the TMS.

The trace and antitrace map method developed here can be expected to
have many applications in the study of one-dimensional aperiodic
systems.

\acknowledgments

XW thanks Shaohua Pan, Guozhen Yang, Changpu Sun, and Hongchen Fu for
their encouragement and kind help. This work has been 
supported by Deutsche Forschungsgemeinschaft.

\appendix
\renewcommand{\theequation}{\Alph{section}.\arabic{equation}}
\section{Relations for unimodular matrices}
\label{sec:rum}

For convenience, we present a collection of relevant identities,
which are used in the construction of the trace and antitrace
maps in Secs.~\ref{sec:tam}, \ref{sec:ass}, and \ref{sec:mem}.

The $n$th power of a unimodular $2\!\times\! 2$ matrix $A$ can be
written as\cite{Kolar90,Wang}
\begin{equation}
A^n=U_n(x_A)A-U_{n-1}(x_A)I,
\label{eq:an}
\end{equation} 
where $I$ is the unit matrix and 
\begin{eqnarray}
U_n(x_A)&=&\frac{\lambda_+^n-\lambda_-^n}{\lambda_+-\lambda_-},\nonumber\\
\lambda_\pm&=&\frac{x_A\pm\sqrt{x_A^2-4}}{2}.\label{eq:un}
\end{eqnarray}
Here $x_A$ and $\lambda_\pm$ denote the trace and the two eigenvalues
of $A$, respectively, and $\lambda_{+}\lambda_{-}=\det A=1$.  The
functions $U_n(x)$ are related to the Chebyshev polynomials of the
second kind $C_n(x)$ by $U_{n}(x)=C_{n-1}(x/2)$. {}From the definition
of the functions $U_n(x)$, it follows
\begin{eqnarray}
U_{-1}(x)&=&-1,\quad U_0(x)=0, \nonumber\\ 
U_1(x)&=&1,\quad U_2(x)=x,\nonumber\\
U_3(x)&=&x^2-1,\quad U_4(x)=x^3-2x, \nonumber\\
U_{n+1}(x)&=&x U_n(x)-U_{n-1}(x),\nonumber\\
U_n^2(x)&=&U_{n+1}(x)U_{n-1}(x)+1.
\label{eq:u}
\end{eqnarray}

In order to study the antitrace maps, we need the following
identity\cite{Dulea90}
\begin{equation}
y_{AB}=x_By_A+x_Ay_B-y_{BA}
\label{eq:yab1}
\end{equation}
for the antitraces of two unimodular $2\!\times\! 2$ matrices $A$ and
$B$.  Now, we briefly prove this identity by introducing an auxiliary
matrix
\begin{equation}
\gamma=\left(\matrix {0&1\cr -1&0} \right),\quad
\gamma^2=-I, \quad
\det(\gamma)=1.
\end{equation}
For the matrix $A$, we have
\begin{equation}
y_A=x_{A\gamma}.
\label{eq:ya}
\end{equation}
Then the antitrace of $AB$ is given by
\begin{equation}
y_{AB}=x_{AB\gamma}=-x_{\gamma A \gamma \gamma B}.
\label{eq:yab2}
\end{equation}
Let $A$, $B$, and $C$ be unimodular matrices. Then\cite{Kolar902}
\begin{eqnarray}
x_{ABAC}&=&x_{AB}x_{AC}+x_{BC}-x_Bx_{C}.
\end{eqnarray}
Applying the above identity to Eq.~(\ref{eq:yab2}) and using
Eq.~(\ref{eq:ya}) again, we obtain Eq.~(\ref{eq:yab1}).

It should be pointed out that Eq.~(\ref{eq:yab1}) is valid for any
pair of $2\!\times\! 2$ matrices, and it follows directly from the
identity
\begin{equation}
AB = (x_{AB}-x_{A}x_{B})I+x_{A}B+x_{B}A-BA
\label{eq:gen}
\end{equation}
which holds for any pair of $2\!\times\! 2$ matrices.  The detailed
proof of this identity can be found in
Ref.~\onlinecite{Avishai}. Here, we only need to consider unimodular
matrices.

For $n=2$, Eq.~(\ref{eq:an}) becomes
\begin{equation}
A^2=x_AA-{I}. 
\label{eq:a2}
\end{equation}
This is the well-known Cayley-Hamilton theorem. 
{}From the theorem, we have 
\begin{eqnarray}
A+A^{-1}&=&x_{ A}I,\nonumber\\
x_{A^{-1}}&=&x_A,\nonumber\\
x_{BA}+x_{BA^{-1}}&=&x_Ax_B,\nonumber\\
y_{BA}+y_{BA^{-1}}&=&x_Ay_B,
\label{eq:amin}
\end{eqnarray}
{}From Eqs.~(\ref{eq:gen}) and (\ref{eq:amin}), we can prove the
following useful relations
\begin{eqnarray}
BAB&=&A-x_{A}I+x_{AB}B=x_{AB}B-A^{-1},\nonumber\\
BA^{-1}B&=&(x_{A}x_{B}-x_{AB})B-A. \label{eq:gen10}
\end{eqnarray}
Finally, from Eqs.~(\ref{eq:an}), (\ref{eq:u}), and (\ref{eq:a2}), we
obtain the following relations
\begin{eqnarray}
x_{A^2}&=&x_A^2-2,\nonumber\\
x_{A^n}&=&U_{n+1}(x_A)-U_{n-1}(x_A),\nonumber\\
y_{A^2}&=&x_Ay_A,\nonumber\\
y_{A^n}&=&U_n(x_A)y_A.\label{eq:gen11}
\end{eqnarray}
This completes our collection of identities. 

\section{Antitrace maps for some metallic mean sequences}
\label{sec:mms}

The trace and antitrace maps for the golden mean and the copper mean
sequences were discussed explicitly in the main part of this
paper. Here, we give the trace and antitrace maps for some other
prominent examples of metallic mean sequences.

{}From Eqs.~(\ref{eq:xxl}), (\ref{eq:yyl}), and (\ref{eq:u}), the
trace and antitrace maps for the silver mean case ($m=2, n=1$) are
obtained as
\begin{eqnarray}
x_{l+1}&=&\frac{x_l}{x_{l-1}}[x_l(x_{l-1}^2-1)-x_{l-2}]-x_{l-1},\\
y_{l+1}&=&\frac{x_l}{x_{l-1}}(y_{l-2}+y_l)+(x_l^2-1)y_{l-1}.
\end{eqnarray}
For the bronze mean sequence ($m=3, n=1$), we find
\begin{eqnarray}
x_{l+1}&=&\frac{x_l^2-1}{x_{l-1}^2-1}
[x_l(x_{l-1}^3-2x_{l-1})-x_{l-2}]
\nonumber\\
&&-x_lx_{l-1},
\\
y_{l+1}&=&\frac{x_l^2-1}{x_{l-1}^2-1}(y_{l-2}+x_{l-1}y_l)
+(x_l^3-2x_l)y_{l-1}. 
\nonumber\\
\end{eqnarray}
Finally, for the nickel mean case ($m=1, n=3$), the result reads
\begin{eqnarray}
x_{l+1}&=&(x_{l-1}^2-1)(x_lx_{l-1}-x_{l-2}^3+3x_{l-2})
\nonumber\\
&&-x_lx_{l-1},\\
y_{l+1}&=&(x_{l-1}^2-1)(x_{l-2}^2-1)y_{l-2}
\nonumber\\
&&+x_l(x_{l-1}^2-1)y_{l-1}-x_{l-1}y_l.
\end{eqnarray}

\newpage

\begin{table}
\narrowtext
\caption{The initial conditions for the trace and antitrace map.
         \label{tab:1}}
\begin{tabular}{ccccc}
      & $m=1$          & $m=2$          & $m=3$          & $m=4$ \\ \hline
$x_1$ & $0$            & $0$            & $0$            & $0$ \\
$x_2$ & $0$            & $-\varrho_{1}$ & $0$            & $\varrho_{1}$ \\
$v_2$ & $-\varrho_{1}$ & $0$            & $\varrho_{1}$  & $0$\\ 
$y_1$ & $2$            & $2$            & $2$            & $2$\\ 
$y_2$ & $\varrho_{1}$  & $0$            & $-\varrho_{1}$ & $0$ \\
$\tilde{w}_2$ 
      & $0$            & $-\varrho_{1}$ & $0$            & $\varrho_{1}$\\ 
\end{tabular}
\end{table}

\begin{table}
\caption{The trace, antitrace, and transmission coefficients for
FC($m$) with $m=2q$. The upper (lower) signs refer to even (odd) values
of $q$.\label{tab:2}}
\begin{tabular}{cccc}
$l$ & $x_l$            & $y_l$ & 
                         $t_l$ \\ \hline 
$1$ & $0$              & $2$   & 
                         $1$\\ 
$2$ & $\pm\varrho_{1}$ & $0$   & 
                         $4/\varrho_{1}^{2}$ \\ 
$3$ & $0$              & $1/\varrho_{m}$   & 
                         $4/\varrho_{m}^{2}$\\ 
$4$ & $\varrho_{1}$    & $0$  & 
                         $4/\varrho_{1}^{2}$\\ 
$5$ & $0$              & $1/\varrho_{2m}$  & 
                         $4/\varrho_{2m}^{2}$\\ 
$6$ & $\pm\varrho_{1}$ & $0$ & 
                         $4/\varrho_{1}^{2}$ \\ 
$7$ & $0$              & $1/\varrho_{3m}$ &  
                         $4/\varrho_{3m}^{2}$ \\ 
\end{tabular}
\end{table}

\begin{table}
\caption{The trace, antitrace, and transmission coefficients for 
FC($m$) with $m=2q+1$. The upper (lower) signs refer to even (odd) values
of $q$.\label{tab:3}}
\begin{tabular}{cccc}
$l$ & $x_l$          & $y_l$                      & 
                       $t_l$ \\ \hline
$1$ & $0$            & $2$                        & 
                       $1$\\ 
$2$ & $0$            & $1/\varrho_{1}$             & 
                       $4/\varrho_{1}^{2}$\\ 
$3$ & $-\varrho_{1}$ & $0$                        & 
                       $4/\varrho_{1}^{2}$ \\ 
$4$ & $0$            & $\mp 1/\varrho_{m+1}$ & 
                       $4/\varrho_{m+1}^{2}$ \\ 
$5$ & $0$            & $1/\varrho_{m}$         &  
                       $4/\varrho_{m}^2$\\ 
$6$ & $\varrho_{1}$  & $0$                        & 
                       $4/\varrho_{1}^{2}$ \\ 
\end{tabular}
\end{table}\clearpage

\end{document}